# Classical Statistics Inherent in a Quantum Density Matrix

## A. K. Rajagopal and R. W. Rendell
### Naval Research Laboratory, Washington, DC 20375

### ABSTRACT

A density matrix formulation of classical bipartite correlations is constructed. This leads to an understanding of the appearance of classical statistical correlations intertwined with the quantum correlations as well as a physical underpinning of these correlations. As a byproduct of this analysis, a physical basis of the classical statistical correlations leading to additive entropy in a bipartite system discussed recently by Tsallis et al emerges as inherent classical spin fluctuations. It is found that in this example, the quantum correlations shrink the region of additivity in phase space.

If the probabilities of a classical statistical model are regarded as the coefficients of a diagonal density matrix, a question arises as to its quantum features. For example, a bipartite system with classical probabilities, $p_{i,j}$, and its associated marginal probabilities $a_i = \sum_j p_{i,j}$, $b_j = \sum_i p_{i,j}$ may be represented by the density matrix in some suitable representation, $\hat{r} = \sum_{ij} p_{ij} |a_{ij}\rangle\langle a_{ij}|$. One normally considers the possible statistical correlations in this system by examining the mutual entropy, $S_1(A:B) = S_1(A) + S_1(B) - S_1(A,B)$. The von Neumann entropy of the bipartite system is $S_1(A,B) = -k \sum_{i,j} p_{i,j} \ln p_{i,j}$ and $S_1(A), S_1(B)$ are the corresponding von Neumann entropies associated with the marginals. If the systems A and B are statistically uncorrelated, i.e. $p_{i,j} = a_i b_j$, then $S_1(A:B) = 0$. If $S_1(A:B) > 0$, then the systems A and B are statistically correlated. In terms of the density matrix, the von Neumann entropy is $S_1(A,B) = -k\, tr(\hat{r} \ln \hat{r})$ and the corresponding definition of mutual entropy is used to measure correlations. This density matrix formulation allows us to explore the quantum aspects (e.g. entanglement) of such statistical models. If $S_1(A:B) \geq 0$, the equality sign implies that $\hat{r}(A,B) = \hat{r}(A) \otimes \hat{r}(B)$ and the inequality sign implies that the density matrix may be a convex sum of products of density matrices of the subsystems A and B. This interpretation has no classical counterpart. To study these features of the density matrix formalism associated with a classical probability model, we explore an interesting classical bipartite statistical model studied by Tsallis, Gell-Mann, and Sato [1], (hereafter referred to as TGS). In the TGS model, the classical correlations are so designed as to make the nonadditive $q$-entropy with index $q$, $S_q(A,B) = k \left[1 - \sum_{ij} p_{i,j}^q \right] / (q-1)$, become additive in the sense that $S_q(A,B) = 2S_q(A) = 2S_q(B)$ for $0 \leq q \leq 1$. For q=1, these *q*-entropies reduce to their von Neumann versions. In the limit q=1, the TGS procedure leads to classically uncorrelated systems for which $S_1(A,B) = 2S_1(A)$ if the systems A and B are equivalent. From the procedure outlined above, this can also be considered in terms of a density matrix, $S_q(A,B) = k(1 - tr(\mathbf{r}^q))/(q-1)$, thus providing a basis for studying the quantum features.

The purpose of the present paper is to explore the TGS model from the density matrix viewpoint to shed light on the quantum correlations that appear in such a procedure as well as the physical nature of the statistical correlations in this model. We focus on the correlations induced by the *q*-additivity as a way to contrast the competition between entanglement and *q*-additivity. We further extend this model to include quantum entanglement in order to examine its consequences both as to the physical and statistical correlations implied by this modification.

We define a quantum bipartite density matrix, $\hat{\mathbf{r}}(A,B)$, corresponding to the classical model of TGS in the computational basis $\{|11\rangle, |10\rangle, |01\rangle, |00\rangle\}$ to represent systems A, B:

$$\hat{\mathbf{r}}(A,B) = (p^2 + \mathbf{k})|11\rangle\langle 11| + [p(1-p) - \mathbf{k}](|10\rangle\langle 10| + |01\rangle\langle 01|) +$$
$$+ [(1-p)^2 + \mathbf{k}]|00\rangle\langle 00| \quad (1)$$

with its marginal density matrices

$$\hat{\mathbf{r}}(A) = p|1\rangle\langle 1| + (1-p)|0\rangle\langle 0|, \quad \hat{\mathbf{r}}(B) = p|1\rangle\langle 1| + (1-p)|0\rangle\langle 0| \quad (2)$$

Here the four bipartite probabilities are $0 \leq p^2 + \mathbf{k}, p(1-p) - \mathbf{k}, (1-p)^2 + \mathbf{k} \leq 1$ and their sum is unity, and the sub-system probabilities are $p, (1-p)$ with $0 \leq p \leq 1$. In this description, statistical correlations are both classical and quantum in nature as will be shown now. Clearly if $\mathbf{k}$ is zero, then the density matrix is $\hat{\mathbf{r}}(A,B) = \hat{\mathbf{r}}(A) \otimes \hat{\mathbf{r}}(B)$ as can be directly verified, and the systems A and B are classically uncorrelated. From eq.(1), it is easily verified that a nonzero value of $\mathbf{k}$ represents a quantum separable state in the sense that the density matrix is a convex sum of products of four density matrices whose coefficients add to unity:

$$\hat{\mathbf{r}}(A,B) = \sum_i w_i \, \hat{\mathbf{r}}_i(A) \otimes \hat{\mathbf{r}}_i(B), \quad 0 \leq w_i \leq 1, \sum_i w_i = 1. \quad (3)$$

We thus note that the density matrix representation of the TGS bipartite probabilities leads to a quantum version with interesting consequences.

The TGS construction gives us non-zero values of $\mathbf{k}$ as a function of p and the nonadditivity parameter, *q* (Fig.1 in TGS), which we here denote as $\mathbf{k} \equiv \mathbf{k}_{q^*}(p)$. If suffices to consider $0 < q < 1$ for the purposes of the discussions in this paper. If $q^* = 1$, $\boldsymbol{k} = 0$ for all p, as expected. Such a nonzero $\boldsymbol{k}$ however represents a classical correlation that renders the *q*-entropy to be additive and gives rise to non-zero mutual entropy defined earlier, which is a measure of classical probability correlation:

$$\begin{aligned}S_1^{Cl}(A:B) &= Tr\boldsymbol{r}(A)\ln \boldsymbol{r}(A) + Tr\boldsymbol{r}(B)\ln \boldsymbol{r}(B) - Tr\boldsymbol{r}(A,B)\ln \boldsymbol{r}(A,B) \\ &= -2\{p\ln p + (1-p)\ln(1-p)\} + (p^2 + \boldsymbol{k})\ln(p^2 + \boldsymbol{k}) \\ &+ ((1-p)^2 + \boldsymbol{k})\ln((1-p)^2 + \boldsymbol{k}) + 2(p(1-p) - \boldsymbol{k})\ln(p(1-p) - \boldsymbol{k}) > 0\end{aligned}$$

(4)

We now construct a quantum density matrix by modifying eq.(1) by including an explicit quantum decoherence term parameterized by a complex number, z:

$$\begin{aligned}\hat{\boldsymbol{r}}_z(A,B) &= (p^2 + \boldsymbol{k})|11\rangle\langle 11| + [p(1-p) - \boldsymbol{k}](|10\rangle\langle 10| + |01\rangle\langle 01|) + \\ &+ [(1-p)^2 + \boldsymbol{k}]|00\rangle\langle 00| + z|10\rangle\langle 01| + z^*|01\rangle\langle 10|\end{aligned}$$

(5)

Note that this has the same marginal density matrices as for the TGS model, given by eq.(2), but has the eigenvalues

$$\{p^2 + \boldsymbol{k}, (p(1-p) - \boldsymbol{k} + |z|), (p(1-p) - \boldsymbol{k} - |z|), ((1-p)^2 + \boldsymbol{k})\} \tag{6}$$

that must all lie between 0 and 1. Note that with $\kappa = 0$ and $|z| < p(1-p)$, $\hat{\boldsymbol{r}}_z(A,B)$ cannot be expressed in the separable form, eq.(3), using the basis chosen in eq.(5). This will be shown to be quantum entangled. This aspect of the quantum correlations in this density matrix may be quantified in terms of the "concurrence" [2], $0 \le C(A,B) \le 1$, which is a measure of bipartite quantum entanglement in the density matrix of eq.(5) but not present in the TGS density matrix, given by eq.(1). For eq.(5), the concurrence is:

$$C(A,B) = 2\max(0, F(z, p, \boldsymbol{k})) \tag{7a}$$

where,

$$\begin{aligned}F(z, p, \boldsymbol{k}) &= 2\left(|z| - \sqrt{(p^2 + \boldsymbol{k})((1-p)^2 + \boldsymbol{k})}\right) \text{ if } (p(1-p) - \boldsymbol{k}) \ge |z| \\ &= 2\left(p(1-p) - \boldsymbol{k} - \sqrt{(p^2 + \boldsymbol{k})((1-p)^2 + \boldsymbol{k})}\right) \text{ if } (p(1-p) - \boldsymbol{k}) \le |z|\end{aligned}$$

(7b)

This quantity is zero if the system is quantum separable as defined in eq.(3). Notice that when z=0, there is still classical correlation as given by the nonzero mutual entropy expressed in eq.(4).

We now consider the TGS procedure for enforcing the additivity of the *q*-entropy of $\hat{r}_z$ and obtain $k, |z|$ as a function of p for various values of the *q*-parameter. For $|z| = 0$ we recover the TGS result; the inclusion of z changes the values of $k$. In Fig. 1(a,b), these are displayed for several values of *q*. Fig.1(a) recaptures the TGS results and is reproduced here to contrast those results when $|z| > 0$ is included. Fig.1(b) shows that for $|z| = 0.2$, the region of the TGS *q*-additivity shrinks to smaller p values as one increases q. This points to the result that quantum entanglement is not favorable to the demand of *q*-additivity of TGS. Fig.2(a) displays the TGS *q*-additivity region as a function of p and $|z|$ for q= 0.5 and in Fig.2(b) the concurrence is shown for the corresponding parameters. The features of Fig.1 can be seen as constant $|z|$ planes in Figs. 2(a) and 3(a). The surface of quantum entanglement when $C(A, B) \neq 0$ is clearly exhibited. In Fig.3(a), the TGS *q*-additivity region for q=0.7 is presented showing a smaller region of ($\kappa, |z|$, p) corresponding to Fig.1(b). Similarly the concurrence in Fig3(b) shows a reduction in quantum entanglement for the same system parameters.

To discern the quantum correlations in the density matrix given by eq.(5), over and above the classical correlations in it, there are several suggestions in the recent literature [3-6]. One of them is the Quantum Deficit [6]:

$$D(A,B) \equiv S_1^{cl}(A:B) - S_1^{qu+cl}(A:B)$$
$$= 2(p(1-p) - k) \ln(p(1-p) - k) -$$
$$- (p(1-p) - k + |z|) \ln(p(1-p) - k + |z|) -$$
$$- (p(1-p) - k - |z|) \ln(p(1-p) - k - |z|)$$

$$\geq 0$$

(8)

This shows that the classical and quantum correlations appear in an inseparable way if we did not know the origins of the respective parameters. When classical correlations are absent, $k = 0$, one has only quantum correlations as long as $p(1-p) > |z|$, as is clear from eqs.(7, 8).

The representation given here of the bipartite state acquires a physical significance if reworked in terms of the standard spin-1/2 language. First of all, we note that the marginal density matrices, eq.(2) may be written as

$$\hat{r}(A) = p|1\rangle\langle 1| + (1-p)|0\rangle\langle 0| \equiv \frac{1}{2}\{\hat{I} + s_z \hat{s}_z\}$$
$$\hat{I} = |1\rangle\langle 1| + |0\rangle\langle 0|, \quad \hat{s}_z = |1\rangle\langle 1| - |0\rangle\langle 0|, \quad s_z = 2p - 1.$$

(9)

and similarly for $\hat{r}(B)$. This means that the subsystems A, B have nonzero mean value of the z-component of pseudo-spin of the TGS model. And the bipartite density matrix is then expressed in the form

$$\hat{r}_z(A,B) = \frac{1}{4}\left\{\begin{array}{l}\hat{I}\otimes\hat{I} + s_z\left(\hat{I}\otimes\hat{s}_z + \hat{s}_z\otimes\hat{I}\right) + C_{zz}\hat{s}_z\otimes\hat{s}_z + \\ + C_{+-}\hat{s}_+\otimes\hat{s}_- + C_{-+}\hat{s}_-\otimes\hat{s}_+\end{array}\right\}$$

where $\hat{s}_+ = |1\rangle\langle 0|$, $\hat{s}_- = |0\rangle\langle 1|$.

(10)

From this we deduce

$$C_{zz} = tr\hat{r}_z(\hat{s}_z\otimes\hat{s}_z) = 4\tilde{k} + (2p-1)^2 = 4k + (tr\hat{r}_z\hat{s}_z)^2 \quad (11)$$

and

$$C_{+-} = tr\hat{r}_z(\hat{s}_+\otimes\hat{s}_-) = z, \quad C_{-+} = tr\hat{r}_z(\hat{s}_-\otimes\hat{s}_+) = z^*. \quad (12)$$

Thus, we find that the pair correlations among z-components of the spin are found to represent the classical correlation, *k* ; the transverse spin fluctuation represents the quantum correlation, z. This physical interpretation shows that the TGS scheme may have a dynamical underpinning, not evident as originally presented [1] or suspected to have possibly no dynamical foundation [7].

In conclusion, we have here explored the TGS construction including the quantum correlation effect in a bipartite density matrix which mimics the classical model of TGS. As a byproduct of this construction, the physical processes that lie beneath the mathematical construction given in TGS are due to the z-components of the pseudo-spin correlations. Also, it is shown in this example that the quantum entanglement works against the TGS *q*-additivity condition.

Thanks are due to Professor Constantino Tsallis for reading an early version of the manuscript and making useful remarks. This work is supported in part by the Office of Naval Research. AKR expresses his thanks to Dr. Andrew Williams of Air Force Research Laboratory, Rome, New York as well as Drs. Joseph Lanza and Kenric Nelson of AFRL/IFEC, Rome, NY for their generous support of this research.

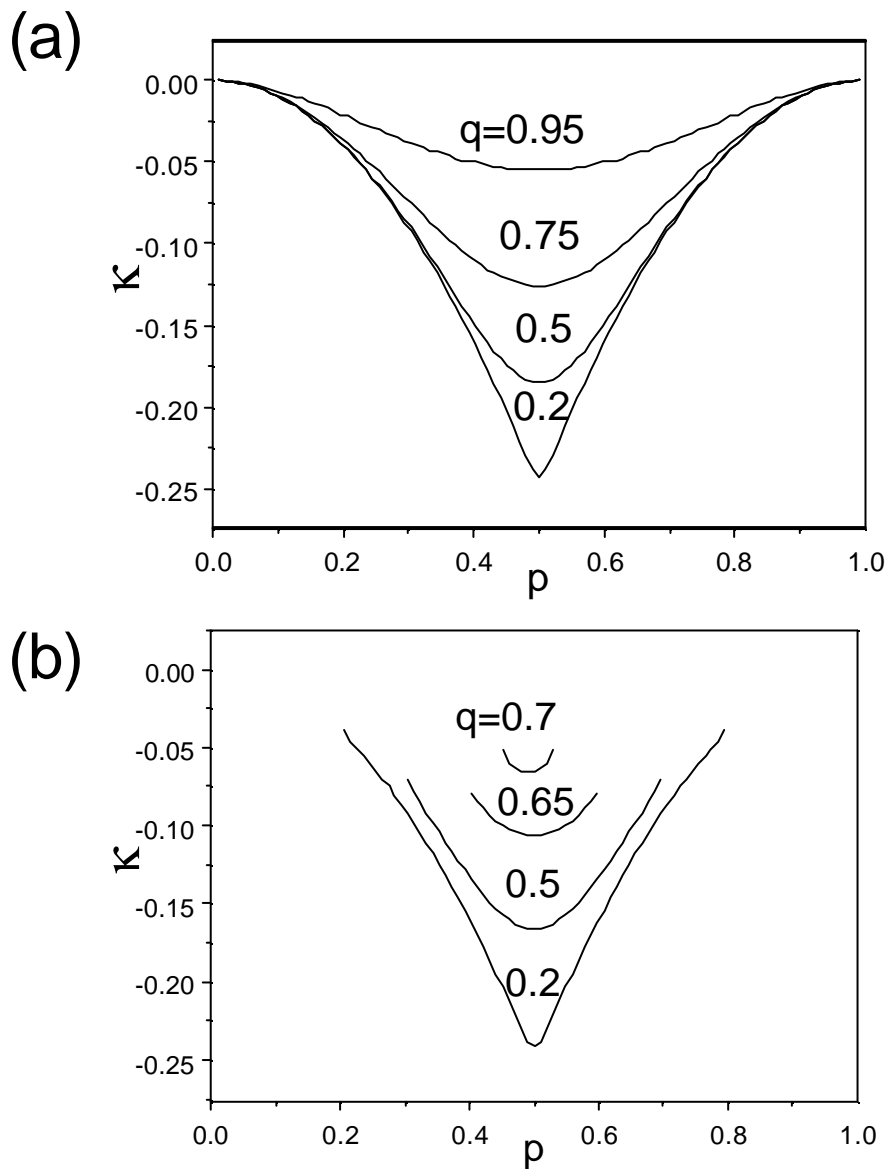

Figure 1. Values of the classical correlation κ obeying the TGS additivity condition as a function of classical probability p at various q values with decoherence values (a) |z| = 0 (TGS model) and (b) |z = 0.2. Note for |z| > 0, the additivity condition has solutions only for restricted ranges of p and q.

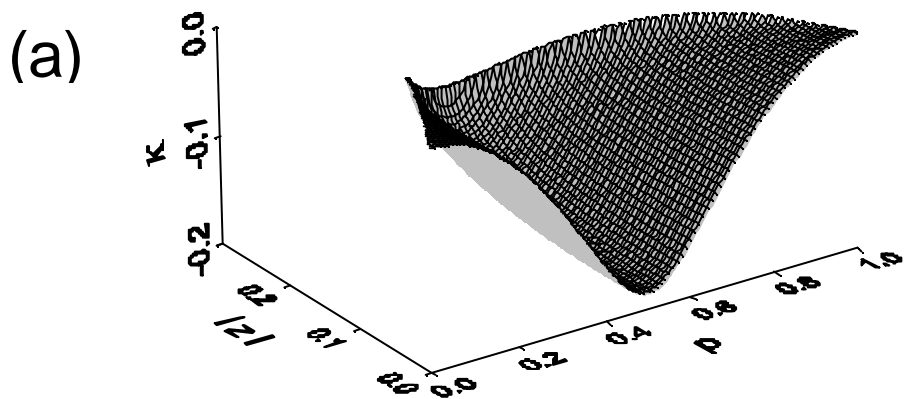

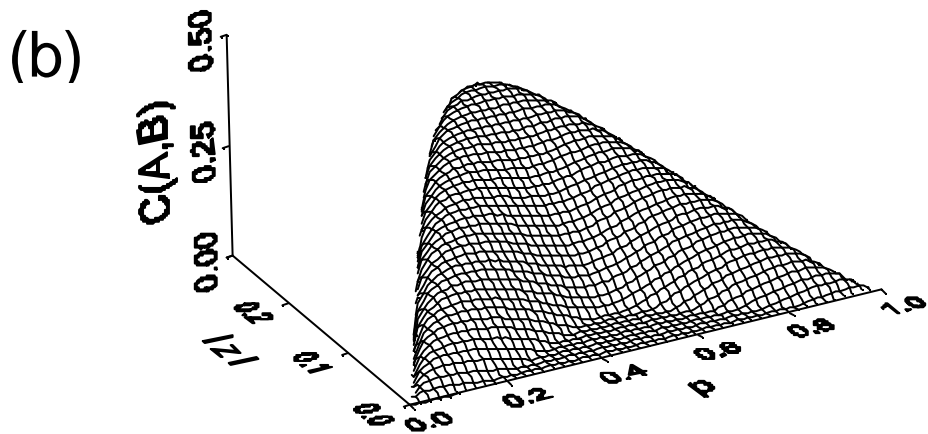

Figure 2. Regions of solutions obeying the TGS additivity condition at q = 0.5 as a function of classical probability p and decoherence z showing (a) classical correlation κ and (b) concurrence C(A,B).

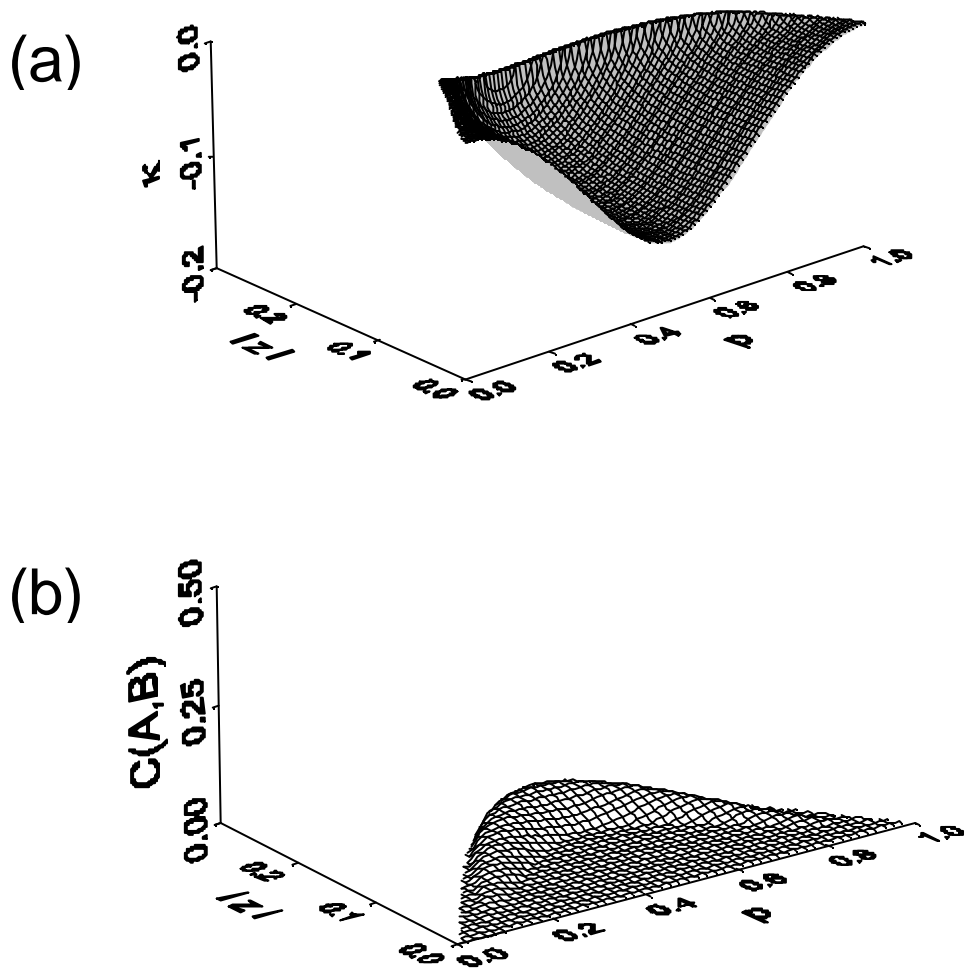

Figure 3. Regions of solutions obeying the TGS additivity condition at q = 0.7 as a function of classical probability p and decoherence z showing (a) classical correlation κ and (b) concurrence C(A,B).